\documentclass[twocolumn,showpacs,preprintnumbers,amsmath,amssymb]{revtex4}

\usepackage[dvips]{graphicx}
\usepackage{dcolumn}
\usepackage{bm}

\begin{document}

\title{Takahashi Integral Equation \\ and High-Temperature Expansion of the Heisenberg Chain} 

\author{Masahiro Shiroishi and Minoru Takahashi}

\affiliation{%
Institute for Solid State Physics, University of Tokyo, \\
Kashiwanoha 5-1-5, Kashiwa, Chiba, 277-8581 Japan 
}%

\date{}

\begin{abstract}
Recently a new integral equation describing the thermodynamics of the 1D Heisenberg model was discovered by Takahashi. 
Using the integral equation we have succeeded in obtaining the high temperature expansion of the specific heat and the magnetic susceptibility up to ${O((J/T)^{100})}$. This is much higher than those obtained so far by the standard methods such as the linked-cluster algorithm.  Our results will be useful to examine various approximation methods to extrapolate the high temperature expansion to the low temperature region. 
\end{abstract}

\maketitle

The spin-1/2 Heisenberg chain is the one of the fundamental models which has been continuously investigated  in the realm of the magnetism in low dimensions. In 1931, Bethe constructed the eigenstates for the isotropic case \cite{Bethe31}, which was later generalized to anisotropic cases such as the ${XXZ}$ and the ${XYZ}$ chain (see \cite{Takahashi99}). The method, which is nowadays called Bethe ansatz method, has provided us vast information on the model. Especially the thermodynamics has been studied through the Thermodynamic Bethe Ansatz (TBA) equations \cite{Takahashi99,Takahashi71,Gaudin71,Takahashi72,Takahashi85, Schlottmann85} and the  Quantum Transfer Matrix method (QTM) \cite{Suzuki87,Koma87,JSuzuki90,Takahashi91n1,Takahashi91n2, Kluemper93,Eggert94,Kluemper98,Kluemper00}. 

Quite recently, one of the authors (MT) has found yet another integral equation \cite{Takahashi01a}, which determines the free-energy of the ${XXZ}$ chain. The integral equation consists of one unknown function and is very simple. We have found that it actually gives the same numerical values for physical quantities as those calculated by the TBA equations and the QTM method. In fact the equation was originally discovered in an attempt to simplify the TBA equations \cite{Takahashi01a}. Soon after that it was derived also from the fusion relation of the QTM \cite{Takahashi01b}. 

In this letter, as one of the applications of this integral equation, we study the high temperature expansion (HTE). As is well known, there is a systematic way to calculate the HTE for any lattice models based on the linked-cluster algorithm \cite{Baker64,Buehler00}. Particularly, in the case of the ${XXX}$ chain, the HTE has been achieved up to ${O((J/T)^{24})}$ \cite{Buehler00}. We have succeeded in obtaining the HTE to much higher order, namely, up to ${O((J/T)^{100})}$ by use of the new integral equation. Below we shall report the result for the isotropic ${XXX}$ model. The generalization to the anisotropic models will be published in a separate paper.

The Hamiltonian of the spin-1/2 Heisenberg ${XXX}$ chain is defined by 
\begin{align}
H&= - J \sum^N_{j=1} \left[ S^x_jS^x_{j+1}+S^y_jS^y_{j+1}+ S^z_jS^z_{j+1} 
 - \frac{1}{4} \right] \nonumber \\
&- 2 h \sum^N_{j=1} S_i^z, \label{XXX}
\end{align}
where ${S_j^{x,y,z}}$ are the local spin-1/2 operators acting on the site ${j}$.We assume the periodic boundary conditions ${S_{N+1} = S_1}$.
Note that in our definition, the coupling constant ${J}$ is positive for the ferromagnetic case and negative for the antiferromagnetic case.  

Takahashi's integral equation for the isotropic ${XXX}$ case is given by
\begin{eqnarray}
u(x)&=& 2 \cosh h/T \nonumber \\
& & + \oint_C \Bigg\{ \frac{1}{x - y - 2 {\rm i}} \exp \left[-\frac{2 J/T}{(y + {\rm i})^2+1} \right]  \nonumber \\
& & + \frac{1}{x - y + 2 {\rm i}} 
\exp \left[-\frac{2 J/T}{(y - {\rm i})^2+1} \right] \Bigg\} \frac{1}{u(y)} \frac{{\rm d} y}{2 \pi \rm i}, \nonumber \\
f &=& - T \ln u(0),
\label{TakahashiEq} 
\end{eqnarray}
where the contour ${C}$ is a loop surrounding the origin in a counterclockwise manner.
The equation can be solved numerically.

In the following we derive the HTE of ${u(x)}$. Actually what we have to do is only to assume ${u(x)}$ in the form 
\begin{align}
& u(x) = \exp \left[ \sum_{n=0}^{\infty} a_n (x) \left(J/T \right)^n \right] \nonumber \\
& = {\rm e}^{a_0 (x)} \Bigg\{ 1+a_1(x) J/T + \left( a_2(x) + \frac{1}{2} a_1(x)^2 \right)  \left(J/T \right)^2  \nonumber \\ 
& \ \ + \left( a_3(x) + a_2(x) a_1(x) + \frac{1}{6} a_1(x)^3 \right)  
\left(J/T \right)^3 + \cdots \Bigg\}. \nonumber \\ \label{ux}
\end{align}
and substitute it into the equation (\ref{TakahashiEq}).
Then by comparing the same order of ${J/T}$ in the LHS and the RHS, we can get the equations, which, to our surprise, determine the functions ${a_n(x)}$ recursively. For example, from the 0-th order, we get immediately
\begin{equation}
a_0(x)= \ln \left( 2 \cosh h/T \right). \label{a0}
\end{equation}
 Similarly by comparing the 1-st order, we have an equation, 
\begin{eqnarray}
a_1(x) &=& \frac{1}{4 \cosh^2 h/T} \Bigg[ \oint_C \Bigg\{ \frac{-2}{y (y+2 {\rm i})(x - y - 2 {\rm i})} 
\nonumber \\
& & \hspace{2cm} + \frac{-2}{y (y - 2 {\rm i})(x - y + 2 {\rm i})} \Bigg\} \frac{{\rm d} y}{2 \pi \rm i} \nonumber  \\ 
& & - \oint_C \Bigg\{ \frac{1}{x - y - 2 {\rm i}} + \frac{1}{x - y + 2 {\rm i}}
 \Bigg\} a_1(y) \frac{{\rm d} y}{2 \pi \rm i} \Bigg] . \nonumber \\
 \label{1st_eq}
\end{eqnarray}
Noting that the second term in the RHS vanishes because the integrand is regular at 
${y=0}$, we can calculate ${a_1(x)}$ explicitly  as  
\begin{equation}
a_1(x) = -\frac{1}{\cosh^2 h/T} \frac{1}{x^2+4}. \label{a1}
\end{equation}

Repeating the similar procedures we can derive each ${a_n(x)}$ successively. For example, we have found 
\begin{eqnarray}
a_2(x) &=& \frac{1}{4 \cosh^2 h/T} \frac{x^2+12}{(x^2+4)^2}  
\nonumber \\
& & - \frac{1}{4 \cosh^4 h/T} \frac{x^2+6}{(x^2+4)^2}, \\
a_3(x) &=& -\frac{1}{24 \cosh^2 h/T} \frac{3 x^4 + 36 x^2 + 160}{(x^2+4)^3} 
\nonumber \\
& & +  \frac{1}{4 \cosh^4 h/T} \frac{x^4 + 11 x^2 + 36}{(x^2+4)^3} 
\nonumber \\
& & - \frac{1}{24 \cosh^6 h/T} \frac{3 x^4 + 30 x^2 + 80}{(x^2+4)^3}.
\end{eqnarray}
In fact, with the help of {\it Mathematica}, we have calculated ${a_n(x)}$ up to ${n=100}$, which provides the HTE for the free energy ${f/T = - \sum_{n=0}^{\infty} a_n(0) (J/T)^{n}}$ up to ${(J/T)^{100}}$. 
Some of the lower terms are given as 
\begin{align}
f/T &= - \ln(2\cosh(h/T)) + \frac{J}{4 T}(1 - B^2) 
\nonumber \\
& - \frac{3J^2}{32 T^2}  \left(1- B^4 \right) 
\nonumber \\
& + \frac{J^3}{192 T^3} (1-B^2)(3 + 7 B^2 + 10 B^4) 
\nonumber \\
& + \frac{5 J^4}{3072 T^4} (1-B^2)(3 - B^2 - 9 B^4 - 21 B^6) 
\nonumber \\
& - \frac{J^5}{5120 T^5}  (1-B^2)(1 + 2 B^2) 
\nonumber \\
& \hspace*{1.5cm} \times (15 - B^2 + 21 B^4 - 63 B^6) 
\nonumber \\
& - \frac{7 J^6}{122880 T^6}  (1-B^2) \nonumber \\
&  \times (3 - 35 B^2 -85 B^4 - 95 B^6 - 30 B^8 
+330 B^{10}) \nonumber \\
& + \dots, \label{HTEfenergy}
\end{align} 
where we have set ${B = \tanh(h/T)}$. From the expansion (\ref{HTEfenergy}), one can get the HTE for other physical quantities. 

For example, we list coefficients of the HTE for the specific heat ${C = -T \frac{\partial^2 f}{\partial T^2}}$ and the magnetic susceptibility ${ \chi = - \frac{\partial^2 f}{\partial h^2}}$ at zero magnetic field in Table \ref{table1} and Table \ref{table2}.   Unfortunately due to the lack of space, we can present the coefficients only up to ${O((J/T)^{51})}$. The coefficients of higer order will be sent on demand to any interested reader.  We remark that the coefficients up to ${O((J/T)^{24})}$ completely coincide with those given in \cite{Buehler00}.( Note the differences of the conventions, ${J \leftrightarrow -J}$ and ${h \leftrightarrow 2h}$. ) The terms with the order larger than ${O((J/T)^{25})}$ are our new results.  It will probably be impossible to get the HTE to such a high order using a conventional method.

\begin{figure}[htbp]
\includegraphics[width=6.5cm]{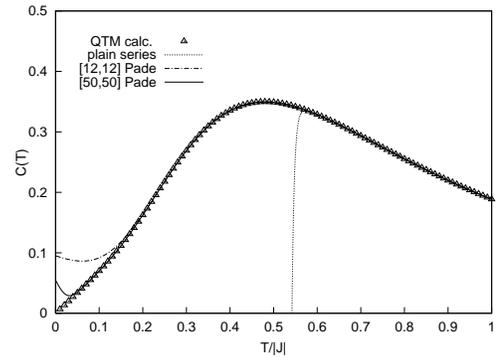} 
\caption{\label{fig:1} Specific heat for the antiferromagnetic 
${XXX}$ chain at ${h=0}$.} 
\end{figure}
\begin{figure}[htbp]
\includegraphics[width=6.5cm]{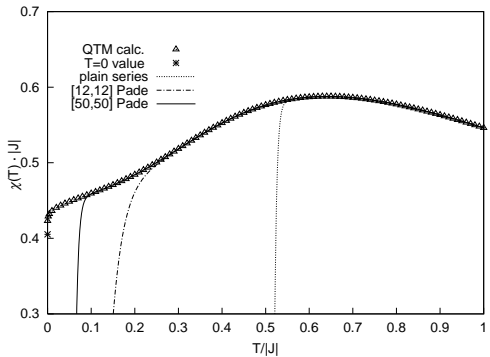}
\caption{\label{fig:2} Magnetic susceptibility for the antiferromagnetic 
${XXX}$ chain at ${h=0}$. ${\chi(0)}$ is taken from \cite{Griffiths64}. } 
\end{figure}
\begin{figure}[htbp]
\includegraphics[width=6.5cm]{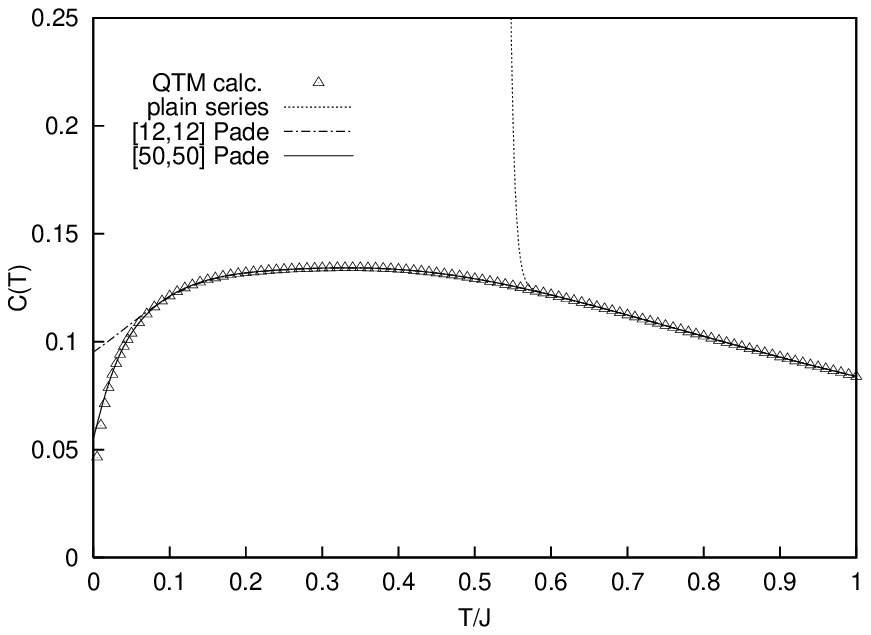}
\caption{\label{fig:3} Specific heat for the ferromagnetic ${XXX}$ 
chain at ${h=0}$.}
\end{figure}
\begin{figure}[htbp]
\includegraphics[width=6.5cm]{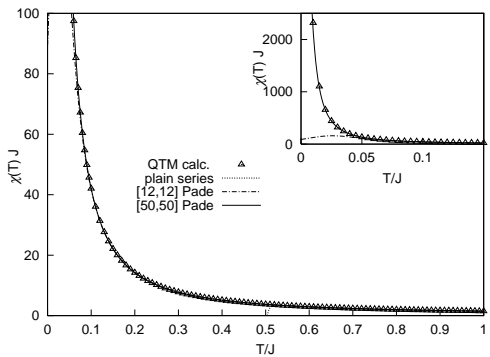}
\caption{\label{fig:4} Magnetic susceptibility for the ferromagnetic 
${XXX}$ chain at ${h=0}$. Note that the plain series of HTE deviates  
from the QTM data around ${T=0.52}$.}
\end{figure}
Usually in order to extrapolate the high temperature expansion series to the low temperature region, some further approximation methods are used.   Actually the original series will not converge in the low temperature region (${T}$ about less than  0.55 ),  because of the existence of the singularities on the complex plane with respect to the inverse temperature. 

Here we have applied the standard Pad\'{e} approximation to our HTE. The results are shown in Fig.~1,...,Fig.~4. Since we have obtained the HTE up to ${O((J/T)^{100})}$, the expressions up to ${[50,50]}$ Pad\'{e} approximant are available. For comparison, we have also plotted the numerical data calculated by the QTM method \cite{Takahashi91n1,Kluemper98}.  (Note that the physical quantities in Fig.~1,...,Fig.~4 were first calculated by the TBA equations. See particularly \cite{Takahashi85,Schlottmann85} for the ferromagnetic case.) From Fig.~1,...,Fig.~4, we find that the ${[50,50]}$ Pad\'{e} approximant show good coincidence with data by the QTM to very low temperature region somewhat like ${T \sim 0.05}$ except for the the magnetic susceptibility for the antiferromagnetic case (Fig.~2.). In that case the logarithmic anomaly around ${T \rightarrow 0}$ is so strong \cite{Eggert94,Kluemper98,Kluemper00} and it probably prevents the good convergence of the Pad\'{e} approximation \cite{Buehler00}.

Apart from the very low temperature region, we have found our higher order Pad\'{e} approximation gives the physical quantities with extremely high precision. For example, we have estimated the peak position of the specific heat and the magnetic susceptibility for the antiferromagnetic case by use of our ${[50,50]}$ Pad\'{e} approximation. The result for the specific heat is 
\begin{align}
C^{\rm max}       &= 0.3497121234553176, \nonumber \\
T^{\rm max}/|J|   &= 0.4802848685890477, \label{Cmax}
\end{align}
and that for magnetic susceptibility is 
\begin{align}
\chi^{\rm max} |J| &= 0.5877051177413559, \nonumber \\
T^{\rm max}/|J|    &= 0.6408510308513831. \label{chimax}
\end{align} 
These values are perfectly identical to the ones in \cite{Kluemper00}, 
where the non-linear integral equations for the QTM were solved numerically very carefully.  (Note that our ${\chi^{\rm max} |J|}$ is four times larger than that in \cite{Kluemper00}, because of different normalization factors.) Note also that the peak position of the magnetic susceptibility (\ref{chimax}) was first determined in \cite{Eggert94}.

For comparison we have also estimated the peak position of the specific heat 
for the ferromagnetic case as 
\begin{align} 
C^{\rm max}   &= 0.1342441913136996, \nonumber \\ 
T^{\rm max}/J &= 0.3326119630964252.
\end{align}

In conclusion, we have shown that the the new integral equation by Takahashi is very useful to calculate analytically the HTE for the 1D Heisenberg model. As far as we know, the HTE of such a high order as ${O((T/J)^{100})}$ has not been achieved for any other models except for free ones.  We have seen that our HTE data together with Pad\'{e} approximation, provide the very accurate numerical data for physical quantities to sufficiently low temperatures. There are several further methods to improve the Pad\'{e} approximation, for example, to allow for the asymptotic expression as ${T \rightarrow 0}$, etc. (see, for example, \cite{Buehler00,Buehler01}). Based on the present results we shall investigate such possibilities in the forthcoming publication.

The authors are thankful to A. Kl\"{u}mper, Y. Nishiyama and K. Sakai for valuable discussions. This work is supported by Grants-in-Aid for the Scientific Research (B) No. 11440103 from the Ministry of Education, Culture, Sports, Science and Technology, Japan.

\begingroup
\squeezetable
\begin{table*}
\caption{\label{table1}Series coefficients ${\alpha_n}$ for the high temperature expansion of the specific heat ${C = \sum_{n} \frac{\alpha_n}{n!} \left(\frac{J}{4 T}\right)^n}$ at ${h=0}$.}
\begin{ruledtabular}
\begin{tabular}{rrrr} 
    ${n}$ & ${\alpha_n}$  & ${n}$ & ${\alpha_n}$ \\ 
0 &  0              & 26 & 484455914465376683487755420408217600 \\ 
1 &  0             & 27 & -1592964364128699671723658807556964352 \\ 
2 &  6              & 28 & -1659222341377723674454893065936371187712 \\ 
3 & -36            & 29 & 53827694891210973745020673240061454581760 \\ 
4 & -360           & 30 & 5090517962961447184851808942927438864711680 \\ 
5 &  7200          & 31 & -388446833192725659973817494776649147157053440 \\ 
6 &  15120         & 32 & -11028658525378274359384407389662010654796546048 \\
7 & -1848672       & 33 & 2255854109806569120380670028755308714386167693312 \\
8 &  11426688      & 34 & -18878702580622989070078793482363993425701807063040 \\
9 &  594846720     & 35 & -11721570087037734701356860480896473609272542720163840 \\
10& -11558004480   & 36 & 527183038642469386328859769728396518893185382125404160 \\
11& -199812856320  & 37 & 52548749252010967993948480669499712309299856992951599104 \\
12&  10106191180800 & 38 & -5382365237582398925074954773487741035075672601159589167104 \\
13&  19376365252608 & 39 & -150281021589219619860159284209265140804955107276364175114240 \\ 
14 & -9289795522775040 & 40 & 44482678475307391762958213932359681737961800852665438128046080 \\
15 & 121944211136778240 & 41 & -670778300712303276022754187872671936481343744506621812675706880  \\
16 & 8791781390116945920 & 42 & -323311185126253530334911142992092649497388429937499549362387156992  \\
17 & -310402124957945954304 & 43 & 19271391500067613736198673193545354611765664770995927250862568636416 \\
18 & -7225535925744106143744 & 44  & 1963797073102024140530884388201619017857423297479642613447074848440320  \\
19 & 643407197363813620776960 & 45 & -261757449501391383349154989821467694962901907072496780634929173522022400 \\
20 & 96147483542540314214400 & 46 & -6715036186134671522475926929150627328836680429076585020863949295740518400 \\
21 & -1279121513829538179364945920 & 47 & 2897640509780835688484069216581936412870887902144153768250804439297575354368 \\
22 & 27962069861743501862336200704 & 48 & 
-67884583842448252705729493380589916284543590592089243545625816663055684075520 \\
23 & 2398518627113966015427501883392 & 49 &  
-278396673545453495106462073222678393434494076172191005087521563009127066632192\
00 \\
24 & -129834725539335848980192847462400 & 50 & 2130333568970965233678580974509426707048585535358286474483865352948915543579033600 \\
25 & -3493877000064415911285457158144000 & 51  & 216827879657653769500650387534438914339017251205042192428944\
424756474567924803174400 \\

\end{tabular}
\end{ruledtabular}
\end{table*}
\begin{table*}
\caption{\label{table2}Series coefficients ${\beta_n}$ for the high temperature expansion of the magnetic susceptibility ${\chi = \frac{1}{T} \sum_{n} \frac{\beta_n}{(n+1)!} \left(\frac{J}{4 T}\right)^n}$ at ${h=0}$.}
\begin{ruledtabular}
\begin{tabular}{rrrr} 
    ${n}$ & ${\beta_n}$  & ${n}$ & ${\beta_n}$ \\ 
0 &  1              & 26 & 108250895627317866042581831619969024 \\ 
1 &  4             & 27 & -17922781862082598948422103131404369920 \\ 
2 &  0              & 28 & 120323704775766862241382488276073447424 \\ 
3 & -64           & 29 & 63726499196979511634418116940394187980800 \\ 
4 &  400           & 30 & -2379574083879736929901931437077028466065408 \\ 
5 &  4032          & 31 & -199921736549129208105099470038165104315334656 \\ 
6 & -89376         & 32 & 17205669691857844357030111005301702510893858816 \\
7 & -163840       & 33 & 408607241970908762765181373729511748361273737216 \\
8 &  26313984     & 34 & -102465082506734431652137262962322034251260126822400 \\
9 & -191334400    & 35 & 1315775281576476974395821022113890916515475892469760 \\
10& -9565698560   & 36 & 545540112377157932817560176143490374271324160383778816 \\
11& 210597986304  & 37 & -27991080178083294022121824180738137056773157996564840448 \\
12&  3486950684672 & 38 & -2463499970265937469283180992880401729855043417164218368000 \\
13&  -203634731188224 & 39 & 284868319106208295442521316317947214008165828662443533926400 \\ 
14 & -127324657152000 & 40 & 6335450833027827824904755798342667046011674072210769010753536 \\
15 & 205019990184689664 & 41 & -2393256391170887534773811023482335548575993905771247989063417856 \\
16 & -3169755454477500416 & 42 & 49322825714830688230041284007303368927055837926052566121716908032 \\
17 & -208763541109969256448 & 43 & 17655809163462320179643661041099923582156280899288659229470449729536 \\
18 & 8342101010835559022592 & 44  & -1192343745774700569487973174622411288186826884326738448519272595456000  \\
19 & 175912858271144581529600 & 45 & -106665300086792728016396630723577790891277245485607114802374328230871040  \\
20 & -18366266410738921187573760 & 46 &  16097399250631032968718675605919227751757635356066434920608362971107688448  \\
21 & 40780317289246872850923520 & 47 & 306091410138596910653186173722480725145635803390769178021236250893761904640 \\
22 & 38668138493195891009425244160 & 48 & -180022014647714649285609157196673479185707562483390012375002362479536\
725032960 
\\
23 & -983734184997038611238624428032 & 49 &  5408449632892687473365326165267541130523650154708192975684794827854640578560000 \\
24 & -75650797544886562610211717120000 & 50 & 17426550003663298365992833617318506452913828517681\
69407986516632569765546101309440  \\
25 & 4622511990582180728868776781545472 & 51  & 1503021976366022386779402749953214978007681637342006785713651741321010\
78803395117056  \\

\end{tabular}
\end{ruledtabular}
\end{table*}
\endgroup


\end{document}